\documentclass[10pt,twocolumn,letterpaper]{article}

\usepackage{cvpr}
\usepackage{times}
\usepackage{epsfig}
\usepackage{graphicx}
\usepackage{amsmath}
\usepackage{amssymb}

\usepackage{tabularx}
\usepackage{algorithm}
\usepackage{algpseudocode}
\usepackage{listings}

\usepackage{multirow}
\usepackage{graphicx}
\usepackage[normalem]{ulem}
\usepackage{pstricks}
\useunder{\uline}{\ul}{}
\algnewcommand\algorithmicforeach{\textbf{for each}}
\algdef{S}[FOR]{ForEach}[1]{\algorithmicforeach\ #1\ \algorithmicdo}
\usepackage[pagebackref=true,breaklinks=true,letterpaper=true,colorlinks,bookmarks=false]{hyperref}

\cvprfinalcopy % *** Uncomment this line for the final submission

\def\cvprPaperID{8} % *** Enter the Workshop Paper ID here

\ifcvprfinal\fi
\begin{document}

% TODO
% remove "we"
%%%%%%

%%%%%%%%% TITLE
%\title{Evolving JPEG XS gains and priorities}
\title{Adapting JPEG XS gains and priorities to tasks and contents}

\author{Benoit Brummer\\
intoPIX\\
Mont-Saint-Guibert, Belgium\\
{\tt\small b.brummer@intopix.com}
\and
%Pascal Pellegrin\\
%intoPIX\\
%Mont-Saint-Guibert, Belgium\\
%{\tt\small p.pellegrin@intopix.com}
%\and
%Antonin Descampe\\
%intoPIX\\
%Mont-Saint-Guibert, Belgium\\
%{\tt\small a.descampe@intopix.com}
%\and
Christophe De Vleeschouwer\\
Universit{\'e} catholique de Louvain\\
Louvain-la-Neuve, Belgium\\
{\tt\small christophe.devleeschouwer@uclouvain.be}
}

\maketitle
%\thispagestyle{empty}

%%%%%%%%% ABSTRACT
\begin{abstract}
Most current research in the domain of image compression focuses solely on achieving state of the art compression ratio, but that is not always usable in today's workflow due to the constraints on computing resources.

Constant market requirements for a low-complexity image codec have led to the recent development and standardization of a lightweight image codec named JPEG XS.

% %In this work we show that JPEG XS can be optimized 
% In this work we optimize JPEG XS gains and priorities using the covariance matrix adaptation evolution strategy, and we show that 
% %JPEG XS can be adapted to specific tasks and contents by optimizing its parameters using evolution strategies. 
% evolution strategies can be used to generate optimized JPEG XS parameters for specific tasks and contents. % such as?
%
%
In this work we show that JPEG XS compression can be adapted to a specific given task and content, such as preserving visual quality on desktop content or maintaining high accuracy in neural network segmentation tasks, by optimizing its gain and priority parameters using the covariance matrix adaptation evolution strategy.

\end{abstract}

%%%%%%%%% BODY TEXT
\section{Introduction}

%Background picture
%   context, relevant work, definition of the problem, motivation
%The JPEG XS codec was standardized in 2019 (ISO/IEC 21122) \cite{isojpegxs-p1} in response to a market demand for lightweight image compression \cite{jpegxsWhitepaper}. 

JPEG has been the most widely used image codec since its introduction in 1992 and more powerful standards such as JPEG2000 have failed to take up consequential market shares due in large part to their added complexity. The JPEG XS codec was standardized in 2019 (ISO/IEC 21122) \cite{isojpegxs-p1} in response to market demand for lightweight image compression \cite{jpegxsWhitepaper}.
Its typical use-cases are to replace uncompressed data flow, whose bandwidth requirement is no longer affordable due to the ever increasing content resolution, and embedded systems, which are bound by the complexity of their integrated circuits and often by their battery capacity. Use-cases often target specific tasks and contents, hence it can be beneficial to optimize an image encoder to account for prior knowledge of its intended application.

Similar optimization work has been produced with other codecs; mainly JPEG, for which quantization tables have been optimized for different tasks using various methods \cite{jpegoptSSIM}\cite{jpegoptSimulatedAnnealing}\cite{jpegoptFeatureDetection}\cite{jpegoptGFWA}, and JPEG2000, which has had its large parameter space explored with the use of genetic algorithms \cite{jpeg2000opt}.%to improve medical imaging quality \cite{jpeg2000opt}.

Akin to the JPEG quantization table \cite{jpeg}, JPEG XS uses a table of gains and priorities for each sub-band. The ISO 21122-1 standard provides a table of gains and priorities (jointly referred to as weights) that maximize peak signal-to-noise ratio (PSNR) \cite[Tab. H.3]{isojpegxs-p1} and states that other values may result in higher quality for certain scenarios. These weights are embedded in the encoded image; the decoder parses them and the choice of optimized weights cannot break compatibility with ISO 21122 compliant decoders.
%Objective
%   purpose

We describe a framework based on the covariance matrix adaptation evolution strategy (CMA-ES) to optimize JPEG XS weights for different metrics and contents. Through this method, decreases of up to 14\% of bits-per-pixel (bpp) at constant quality are achieved on desktop content and up to 59\% at constant accuracy on a semantic segmentation task.
\section{Background}\label{sec:Background}
%This section provides a brief overview of the JPEG XS codec (\ref{sec:JPEGXS}) and of CMA-ES (\ref{sec:CMAES}).% as is relevant to the weight optimization process.
\subsection{JPEG XS}\label{sec:JPEGXS}

The JPEG XS coding scheme \cite{isojpegxs-p1} first applies a reversible color conversion from the RGB color space to YC$_b$C$_r$. The signal is then decomposed into a set of sub-bands, through an asymmetric 2D multilevel discrete wavelet transformation (DWT) \cite{dwt}.
Wavelet coefficients are split into precincts containing coefficients from all sub-bands that make up a spatial region of the image (usually a number of horizontal lines), and are encoded as 16-bit integers prior to quantization.
The gains (G$_\text{b}$) and priorities (P$_\text{b}$) table makes up the weights, a pair of values defined for each sub-band (b), which are stored in the encoded picture header. The precinct quantization ($Q_\text{p}$) and precinct refinement ($R_\text{p}$) are computed for each precinct (p) based on the bit budget available for that precinct. The truncation position ($T_\text{b,p}$) determines how many least significant bits are discarded from all wavelet coefficients of a given sub-band within a precinct;
$T_\text{b,p}$ is calculated from G$_\text{b}$, P$_\text{b}$, $Q_\text{p}$, and $R_\text{p}$ as follow: ${T}_{\text{p},\text{b}}=Q_\text{p}-\text{G}_\text{b}+r$ where $r$ is $1$ for $\text{P}_\text{b}<R_\text{p}$ and $0$ otherwise. Eventually $T_{\text{p},\text{b}}$ is then clamped to the valid range, i.e. [0;15]. Bitplane counts (i.e. number of non-zero bitplanes in a group of four coefficients) are then entropy-coded and packed along with the truncated coefficients values, their signs, and $Q_\text{p}$ and $R_\text{p}$ values.

\subsection{CMA-ES}\label{sec:CMAES}

CMA-ES is an evolutionary algorithm which performs derivative-free numerical optimization of non-linear and non-convex functions, using a population of candidate solutions whose mean and covariance matrix are updated for each generation using the most fit individuals. \cite{cmaes}\cite{cmatutorial}

Rios and Sahinidis \cite{cmaes-comp} have extensively reviewed 22 derivative-free optimization algorithms over a wide range of problems. % In this work, we are specifically interested in non-convex non-smooth problems, and the number of function evaluations is not particularly important because this optimization process is typically performed only once. In \cite[Figure~12]{cmaes-comp}, the different algorithms are ranked by the fraction of problems for which they found the best solution after up to 2500 function evaluations. This shows CMA-ES coming up third place %behind TOMLAB/MULTIMIN and TOMLAB/LGO, and ahead of PSWARM and MCS.
They show that CMA-ES outperforms other algorithms by $\sim$200\% when measuring the fraction of non-convex non-smooth problems with 10-30 variables solved from a near-optimal solution \cite[Figure~32]{cmaes-comp}. The optimization of JPEG XS High profile weights fits this use-case, given there are 30 variables whose initial values have already been optimized to maximize PSNR, and the number of function evaluations is of little importance since this optimization is performed only once. % trimmable (eg: rm competition)
The pycma library \cite{pycma} provides a well-tested implementation of CMA-ES.

%-------------------------------------------------------------------------
\section{Experiments}\label{sec:Experiments}
The optimization method described in \ref{sec:Method} introduces the CMA-ES optimizer and its pycma implementation, details the use of a JPEG XS codec, and lists the datasets and metrics used. Results are given in \ref{sec:Results}.
%This section details the optimization methods in \autoref{sec:Method}, then lists the results in \autoref{sec:Results}. 
\subsection{Method}\label{sec:Method}
\smallskip\noindent\textbf{CMA-ES implementation}\smallskip

% pycma
We use the pycma \cite{pycma} CMA-ES implementation. The only required parameters are an initial solution $X_0$ and the initial standard deviation $\sigma_0$.
The gains defined in \cite[Tab. H.3]{isojpegxs-p1} are used for $X_0$ and their standard deviation is computed for $\sigma_0$. Population size is set to 14 by pycma based on the number of variables. Gains are optimized as floating-point values; their truncated integer representation is used as the gains given to the encoder and the fractional parts are ranked in descending order which is used as priorities. This matches the relative importance of priorities in Section \ref{sec:JPEGXS}, where a bit of precision is added to bands whose priority $\text{P}_\text{b}$ is smaller than the precinct refinement threshold $R_\text{p}$.

\medskip\noindent\textbf{JPEG XS implementation}\smallskip

% JPEG XS
JPEG XS High profile \cite[Tab. 2]{jpegxsWhitepaper} is optimized using TICO-XSM, the JPEG XS reference software \cite{isojpegxs-p5} provided by intoPIX \cite{ticoxs}. Each image is encoded and decoded with a given set of weights and the average loss (1-MS-SSIM, -PSNR, or the prediction error) on the image set is used as the fitness value. A single image is encoded and decoded using the \verb|tco_enc_dec| command with a given target bpp and a configuration file containing the candidate weights.
% using the command ``\verb|tco_enc_dec -p 11 -b [bpp]|
% \verb|-C [config_file.ini] [in.png] [out.png]|'', where \verb|config_file.ini| has the following syntax:
% \begin{verbatim}
% [levels]
% gains_choice = gains_choice_manual
% lvl_gains = {[gains]}
% lvl_priorities = {[priorities]}
% \end{verbatim}.
The gains and priorities are listed in the configuration file in the following order (deepest level first): LL$_{5,2}$, HL$_{5,2}$, HL$_{4,2}$, HL$_{3,2}$, HL$_{2,2}$, LH$_{2,2}$, HH$_{2,2}$, HL$_{1,1}$, LH$_{1,1}$, HH$_{1,1}$ where H denotes high-pass filtering and L is low-pass filtering. The list is repeated for each component (Y, C$_b$, C$_r$). This differs from \cite[Tab. H.3]{isojpegxs-p1}, where the three components are alternating for each value.

% Data
\medskip\noindent\textbf{Human visual system optimization}\smallskip

The training data used to generate weights optimized for image quality is a random subset of 240 featured pictures from Wikimedia Commons \cite{featuredpictures}, randomly scaled between 0.25 and 1.00 of their original size and cropped to match the Kodak Image Dataset \cite{kodak} resolution of 768x512. A different subset of 240 featured pictures and the 24-pictures Kodak Image Dataset are used for testing. A set of 200 screenshots has been collected on Wikimedia Commons for synthetic desktop content optimization, ensuring variability in the software, operating systems, fonts, and tasks displayed. The desktop content has a resolution of 1920x1080 and is split into two 100-picture subsets for training and testing. These images are optimized for MS-SSIM and PSNR.
%JPEG XS are optimized for an AI task 
%In order 
%AI weights are optimized 
%

\medskip\noindent\textbf{Computer vision optimization}\smallskip

Optimized weights are generated for an AI-based computer vision task to minimize prediction error incurred from compression artifacts. This employs the Cityscapes dataset, consisting of street scenes captured with an automotive camera in fifty different cities with pixel-level annotations covering thirty classes. \cite{cityscapes} The data is split between a ``train'' set used to train a convolutional neural network (CNN) that performs pixel-level semantic labeling on uncompressed content, and a ``val'' set used to test the accuracy of the trained CNN using the intersection-over-union ($\text{IoU}={\text{TP}\over{\text{TP}+\text{FP}+\text{FN}}}$) metric. % A ``test'' set is also provided ...
%A HarDNet \cite{hardnet} convolutional neural network (CNN) is trained to perform pixel-level semantic labeling on the uncompressed Cityscapes dataset \cite{cityscapes}, and its baseline accuracy is measured as the intersection over union (IoU) on validation data. 
%. The performance of an AI task is measured as the intersection over union (IoU) score obtained by a HarDNet \cite{hardnet} semantic segmentation convolutional neural network (CNN) trained to perform pixel-level semantic labeling on the Cityscapes dataset \cite{cityscapes}.
The HarDNet architecture is well suited for this optimization task because it performs fast inference without forgoing analysis accuracy \cite{hardnet}, enabling faster (and thus more) parameter evaluations in CMA-ES. 
%(move cityscapes before hardnet?)

The ``train'' data cannot be used to evaluate IoU performance because the model has already seen it during training; therefore, we optimize the 500-image "val" data. This is split into three folds: two cities used as training data and one city used as testing data.
%(among Frankfurt, Lindau, and Munster). <- I don't think the choice of city is necessary unless you want to describe why you picked each?
The weighted average over three folds is based on the number of images in each test set. The MS-SSIM metric is also optimized on the Cityscapes dataset; the ``train'' and "train\_extra" data can be reused, making two 100-images subsets for training and testing.
%, each out of twenty of the 41 different cities. <- is this important enouch to mention?

Similarly to the human visual system (HVS) targeted optimization, AI optimization is performed by encoding and decoding the whole set of images to be evaluated, computing the average IoU metric, and providing its result to the optimizer as the fitness value of the current set of weights.

% mention validation imbalance
\medskip\noindent\textbf{Hardware and complexity}\smallskip

The weights optimization process is parallelized using all available CPU threads, each encoding an image. Image analyses such as semantic segmentation and MS-SSIM evaluation are computed faster on a GPU. Two or more optimization jobs are run in parallel, making constant use of all available computing resources. 27 weight configurations are extensively optimized and results are obtained in approximately three weeks from a total of 78 CPU threads.

\phantomsection\label{par:interpolation}
Each training process is repeated at 1.00, 3.00, and 5.00 bpp. 4000 function evaluations are performed for each visual optimization and 1500 for each AI task. In addition, the AI task optimization is repeated over three folds and different bitrates are interpolated to produce Figure \ref{fig:iou}. The interpolation is carried out by computing the fitness value of the top-10 weights of the closest optimized bitrate at each interpolated bitrate, as well as performing a smaller 150 function evaluation optimization at the interpolated bitrates.

\subsection{Results}\label{sec:Results}

\noindent\textbf{Human visual system results}\smallskip

Results of the HVS weights optimization are summarized in Table \ref{tab:msssim}. Weights have been optimized for different metrics and contents and tested with the MS-SSIM and PSNR metrics. Table \ref{tab:msssim} shows that MS-SSIM is consistently improved through the use of optimized weights, even when using different content classes. Improvement is especially considerable at low bitrate, as the standard weights consume 12\% to 18\% more bitrate to achieve the MS-SSIM index obtained with 1.00 bpp using optimized weights. These results are close to those obtained using the ``visual'' weights provided in the reference software. Improvements obtained with PSNR weights are negligible; the standard weights are already optimized for this metric and content-specific optimizations bring no significant improvement. Weights optimized for an AI task (IoU on Cityscapes) perform poorly when measured against PSNR or MS-SSIM, even on the same dataset.% this highlights the potential usefulness of AI-specific optimization. %A visual comparison of the MS-SSIM optimized weights is shown on Figure \ref{fig:vismsssim}
%Add bpp gain at each bpp from MS-SSIM to ISO PSNR
% \usepackage{multirow}
% \usepackage{graphicx}
% \usepackage[normalem]{ulem}
\begin{table*}[]
%\begin{center}
\centering
\caption{Results obtained with weights optimized for different contents and visual metrics (optimization shown in the left two columns). The test metrics, content, and bpp are shown on top. The best weight for each test is marked as such in bold. The last two columns show the bpp percentage increase needed by standard (or visual \cite{isojpegxs-p5}) weights to match the score of the optimized (underlined) weights.} %The last two rows shows the bpp needed with standard (or visual \cite{isojpegxs-p5}) weights to match the score of the optimized (underlined) weights.}
%\vspace{1pt}
\resizebox{\textwidth}{!}{%

\begin{tabular}{ll|lll|lll|lll|l|lll}
%\hline
\multicolumn{2}{l}{\multirow{2}{*}{}}                       & \multicolumn{9}{|l|}{{\ul MS-SSIM}}                                                                                                                                       & \multicolumn{4}{l}{{\ul PSNR}}                                \\% \cline{3-15} 
\multicolumn{2}{l}{}                                        & \multicolumn{3}{|l|}{{\ul FeaturedPictures}}            & \multicolumn{3}{l|}{{\ul kodak}}                       & \multicolumn{3}{l|}{{\ul screenshots}}  & \multicolumn{1}{l|}{{\ul kodak}}                    & \multicolumn{3}{l}{{\ul screenshots}}              \\
{\ul weights dataset}                   & {\ul weight metric} & 1.00 bpp          & 3.00 bpp          & 5.00 bpp          & 1.00 bpp          & 3.00 bpp          & 5.00 bpp          & 1.00 bpp          & 3.00 bpp          & 5.00 bpp          & 3.00 bpp         & 1.00 bpp         & 3.00 bpp         & 5.00 bpp         \\ \hline
\multirow{2}{*}{Cityscapes}       & IoU           & 0.96002          & 0.98799          & 0.99240          & 0.95979          & 0.99299          & 0.99713          & 0.98344          & 0.99822          & 0.99946 &38.451         & 30.337          & 50.302          & 59.631          \\
                                        & MS-SSIM       & 0.96631          & 0.99121          & 0.99440          & 0.96711          & 0.99574          & 0.99888          & 0.98780          & 0.99897          & \textbf{0.99974} & 39.183 & 31.541          & 50.976          & 62.450          \\ \hline
\multirow{2}{*}{Desktop}          & MS-SSIM       & 0.96632          & 0.99128          & 0.99441          & 0.96644          & 0.99571          & 0.99887          & {\ul \textbf{0.99216}} & {\ul 0.99898}          & {\ul \textbf{0.99974}}  & 39.573   & 36.393          & 52.289          & 62.612          \\
                                        & PSNR          & 0.95965          & 0.99026          & 0.99406          & 0.95907          & 0.99516          & 0.99869          & 0.98942          & 0.99869          & 0.99967   & 40.959  & {\ul \textbf{38.228}} & {\ul \textbf{53.584}} & {\ul \textbf{64.124}} \\ \hline
\multirow{2}{*}{FeaturedPictures} & MS-SSIM       & {\ul \textbf{0.96699}} & {\ul \textbf{0.99133}} & {\ul \textbf{0.99442}} & {\ul \textbf{0.96717}} & {\ul \textbf{0.99580}} & {\ul \textbf{0.99889}} & 0.99176          & \textbf{0.99906} & 0.99973       & 39.772         & 35.932          & 52.273          & 63.091          \\
                                        & PSNR          & 0.96133          & 0.99037          & 0.99415          & 0.96098          & 0.99516          & 0.99876          & 0.98971          & 0.99886          & 0.99970 & {\ul \textbf{41.014}}  & 37.725          & 53.482          & 64.087          \\ \hline
{JPEG-XSM visual}                         & {Human}         & 0.96547          & 0.99079          & 0.99416          & 0.96569          & 0.99556          & 0.99877          & 0.99137          & 0.99891          & 0.99970     & 39.324          & 35.937          & 50.477          & 60.651          \\ 
ISO 21122 std.                          & PSNR          & 0.96060          & 0.99038          & 0.99416          & 0.96009          & 0.99522          & 0.99876          & 0.98957          & 0.99885          & 0.99970      & 40.999              & 38.147          & 53.569          & 64.085          \\ \hline\hline
\multicolumn{2}{l|}{\% bpp improvement over visual weights} & 4.00 & 6.00 & 6.80 & 3.00 & 2.77 & 3.40 & 6.00 & 7.00 & 5.00 & 18.7 & 23.0 & 19.7 & 15.8 \\ %\hline
\multicolumn{2}{l|}{\% bpp improvement over std. weights} & 14.0 & 9.67 & 6.4 & 12.0 & 5.33 & 3.80 & 18.0 & 5.33 & 4.00 & 3.01 & 1.00 & 0.33 & 0.2 \\ %\hline
%\multicolumn{2}{l|}{min. bpp needed using visual weights} & 1.04 & 3.18 & 5.34 & 1.03 & 3.08 & 5.17 & 1.06 & 3.21 & 5.25 & 3.56 & 1.23 & 3.59 & 5.79 \\ %\hline
%\multicolumn{2}{l|}{min. bpp needed using std. weights} & 1.14 & 3.29 & 5.32 & 1.12 & 3.16 & 5.19 & 1.18 & 3.16 & 5.20 & 3.01 & 1.01 & 3.01 & 5.01 \\ %\hline
\end{tabular}%
}
%\end{center}
%\vspace{-1pt}

\label{tab:msssim}
\end{table*}
%\vspace{-1pt}
\begin{figure*}
  \centering
    \includegraphics[width=\textwidth]{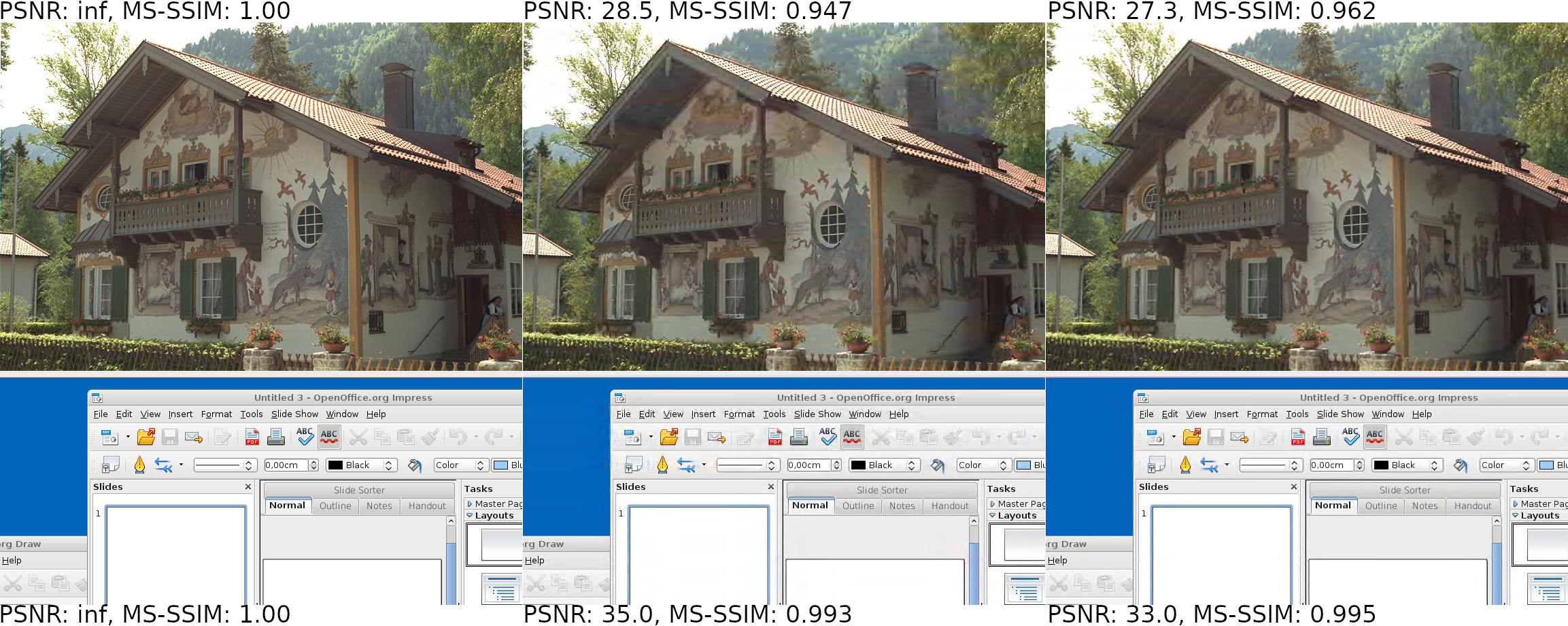}
    \caption{Visual comparison of ``Featured Pictures'' and ``Desktop'' MS-SSIM weights at 1.00 bpp. Left: uncompressed image, middle: compressed with ISO 21122 PSNR weights, right: compressed with MS-SSIM optimized weights (ours), top: kodak image 24, bottom: Tails 0.12 office screenshot.}
    \label{fig:vismsssim}
  %\end{center}
  %%\vspace{-0.2cm} 
\end{figure*}
Figure \ref{fig:vismsssim} shows a visual comparison with weights optimized for the MS-SSIM metric on ``Featured Pictures'' and ``Desktop'' content at 1.00 bpp. MS-SSIM weights tend to yield a consistent level of detail for a globally acceptable visual quality at low bitrate, whereas PSNR-optimized weights tend to show a mix of sharper areas and bleeding artifacts.

% \begin{figure}
%   \begin{center}
%     \includegraphics[width=\linewidth]{../graphics/ai.jpg}
%     \caption{Semantic segmentation on an image from the Cityscapes dataset (Lindau). top-left: ground-truth, bottom-left: segmentation on uncompressed image, top-right: segmentation on image compressed at 1 bpp with ISO 21122 PSNR weights, bottom-right: segmentation on image compressed at 1 bpp with optimized weights}
%     \label{fig:ai}
%   \end{center}
% \end{figure}

% \usepackage{multirow}
% \usepackage{graphicx}
% \usepackage[normalem]{ulem}

\medskip\noindent\textbf{AI results}\smallskip

\begin{figure}
%\vspace{-8pt} 
  \centering
    \includegraphics[width=\linewidth]{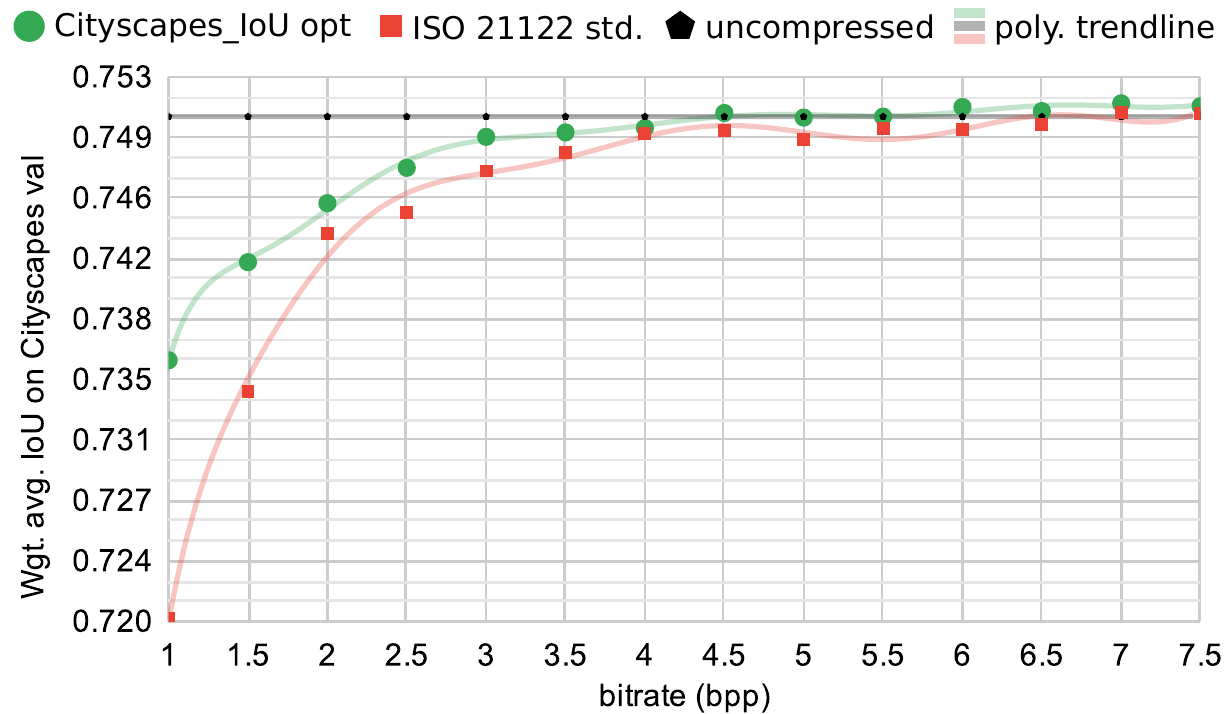}
  \caption{IoU/bitrate performance on Cityscape validation set. JPEG XS standard weights compared to optimized weights (weighted average over three folds). Optimization is performed at 1.00, 3.00, and 5.00 bpp, other bitrates are interpolated as described in section \ref{par:interpolation}.}
  \label{fig:iou}
\vspace{-3pt} 
%\vspace{15pt}
\end{figure}

Table \ref{tab:iou} shows the performance improvements from weights optimized on a semantic segmentation task on the Cityscapes dataset, as well as content-specific MS-SSIM weights. The fully optimized weights achieve a 33.3\% to 59\% reduction in required bpp, meanwhile the MS-SSIM weights do not translate to performance gains on the IoU metric. Figure \ref{fig:iou} shows the gains over many more bitrates, by testing the standard weights every 0.50 bpp and interpolating the optimized weights. The IoU score obtained on uncompressed content is 0.7506. This number is sometimes exceeded when performing inference on compressed content; the highest scores obtained are 0.7514 with optimized weights and 0.7508 with standard weights at 7.00 bpp. 

\begin{table}[]
%\vspace{-0.2cm} 
\centering
\caption{Average IoU obtained over 3 folds using weights optimized for semantic segmentation on Cityscapes. Also shown are weights optimized for MS-SSIM on Cityscapes, and the MS-SSIM index obtained at 3.00 bpp (rightmost column).}
\vspace{2pt}
\resizebox{\linewidth}{!}{%
\begin{tabular}{ll|lll|l}
\multicolumn{2}{l|}{{\ul }}                    & {\ul IoU}              & {\ul }                 & {\ul }                 & {\ul MS-SSIM}                      \\
weight dataset               & weight metric  & 1.00 bpp                  & 3.00 bpp                  & 5.00 bpp                    & 3.00 bpp                           \\ \hline
\multirow{2}{*}{Cityscapes}  & IoU            & {\ul \textbf{0.73583}} & {\ul \textbf{0.74937}} & {\ul \textbf{0.75054}}    & 0.99791                      \\
                             & MS-SSIM        & 0.72882                & 0.74707                & 0.74971                & {\ul \textbf{0.99876}}  \\ \hline
JPEG-XSM visual & Human & 0.73164 & 0.74742 & 0.74990 & 0.99862 \\
ISO 21122 std.                   & PSNR           & 0.72018                & 0.74731                & 0.74920                             & 0.99857                               \\ \hline\hline
%\multicolumn{2}{l|}{min. bpp needed using std. weights} & 1.59 & 4.00 & 6.74       & 3.30                       \\
\multicolumn{2}{l|}{\% bpp improvement over std. weights} & 59.0 & 33.3 & 34.8       & 10.0                       \\
\end{tabular}%
}
%\vspace{0.1px}
\label{tab:iou}
\vspace{-3pt} 
\end{table}

Our results indicate that JPEG XS compression at a high enough bitrate does not appear to be detrimental to the performance of a semantic segmentation model trained on uncompressed content (starting at 4.25 bpp with optimized weights, or 6.90 bpp with standard weights), and it may even be beneficial in some instances, as the model performed better at 7.00 bpp than with uncompressed content.

\medskip %TODO check impact
The metric used for fitness evaluation appears to be much more effective than the type of data provided. This is seen with synthetic ``Desktop'' content optimized weights providing an MS-SSIM index on natural content close to that obtained with the weights optimized on featured pictures, PSNR weights bringing little to no benefit regardless of the content, and the MS-SSIM and IoU optimized weights not benefiting each other on the same content. Moreover, optimizing for different bitrates appears to be beneficial as different sets of weights are generated for each optimized bitrate.%rmed different sets of weights

\section{Conclusion}\label{sec:Conclusion}

% compression may be advantageous?

CMA-ES is a black-box optimization method that can be used to optimize JPEG XS quantization parameters for a given task and content type. %This optimization process would be unnecessarily challenging to implement using the backpropagation algorithm because hard quantization is not differentiable, and
Such evolutionary algorithms work well with the small number of quantization parameters present in JPEG XS and CMA-ES provides a nearly parameter-free method for task-specific optimization of JPEG XS image compression.
%The small number of quantization parameters in JPEG XS (30 in High profile) makes genetic algorithms a good fit to solve this optimization problem%, compared to back-propagation methods which 

A relatively small training set is used. This is necessary because each image needs to be encoded and decoded to test the given weights and scoring must remain consistent for each evaluation. Overfitting does not appear to be an issue even when the training and test sets are small and distinct, likely because the small number of weights do not provide the capacity to overfit.

PSNR weights optimized with CMA-ES do not vary significantly from those defined in the ISO standard, which were calculated to achieve maximum PSNR, regardless of the content type and bitrate used. MS-SSIM optimized weights are more beneficial; a bitrate reduction of 0.12 to 0.32 bpp (3.8\% to 18\%) is observed between 1.00 and 5.00 bpp (depending on the type of content) and the gain in MS-SSIM index translates to a higher perceptual quality.

The weights optimized for an AI-analysis task show the versatility of this method, as they adapt to a very specific task and content and provide a more notable reduction in required bitrate at constant accuracy (33.3\% to 59\% at the optimized bitrates). The fitness function can be tuned to fit any objective function. For example, one could combine the MS-SSIM index and an AI metric to create hybrid human-machine weights.%, a text-centered application could base the fitness function on the performance of an optical character recognition system, and a mask can be applied prior to evaluation to ensure that only specific parts of the images are optimized for.

This search for optimized weights is by no means exhaustive as there are more use-cases to optimize for than could possibly be listed in this work. Even so, it presents a simple and effective method to optimize JPEG XS quantization parameters for any specific task. Furthermore, this method may be used to optimize JPEG XS image compression for AI image analysis models which would themselves be prohibitively expensive (or impossible) to fine-tune for compressed content.

% TODO mention metric more important than weight

%This weight optimization is by no means exhaustive 
% method does not require fine-tuning a model for compression (although it could be done as an iterative process)

% search for weights is by no means and aim is not to give a definitive set of weights but offers a method to generate weights adapted to the application and content at hand without compromising interoperability

% benefit of per bpp weights. content brings an improvement but not huge. psnr is useless (slight gain with same content class, slight loss otherwise)

% the weights found differ at different bpp 

%
%
\section{Acknowledgements}\label{sec:Acknowledgements}
I would like to thank Antonin Descampe for his help in formalizing the JPEG XS background section, and Pascal Pellegrin for helping me understand the JPEG XS codec and analyzing the resulting weights. This research has been funded by the Walloon Region.

{\small
\bibliographystyle{ieeetr}

\bibliography{Evolving_JPEG_XS_gains_and_priorities}
}

\end{document}